\documentstyle[twoside,fleqn,espcrc2]{article}

\def\mr{\rm}	

\def\slsh#1{\rlap{$\kern0.125em /$}#1}

\newcommand{\cD}{{\cal D}}

\def\Order{{\mr O}}

\newcommand{\half}{\mbox{\small $\frac{1}{2}$}}

\def\imag{\mathop{\mr Im}}

\def\real{\mathop{\mr Re}}

\def\zahlen{{\sf Z\kern-0.4em Z}}

\title{Desperate{\rm ly Seeking} Chiral Fermions
\hfill {\normalsize NORDITA--95/85~P}}

\author{A.S. Kronfeld\address{
Nordita, Blegdamsvej 17, DK-2100 K\o benhavn \O, DENMARK}%
\thanks{permanent address:
Theoretical Physics Group, Fermi National Accelerator Laboratory,
P.O. Box 500, Batavia, IL 60510-0500, U.S.A.}%
\thanks{Talk given at {\em Lattice '95}, 11--15 July 1995, Melbourne,
Australia}
\hfill hep-lat/9601008
}

\begin{document}

\begin{abstract}
Chiral fermions can (presumably) be constructed by introducing two
regulators, one for the gauge fields (e.g. a lattice),
and another for the fermion functional integrals in a fixed (regulated)
gauge field.
This talk
discusses cutoff effects arising from the regulator of the fermions.
\end{abstract}

\maketitle

\section{INTRODUCTION}
It has proven difficult to couple chiral fermions to lattice gauge
fields~\cite{Sha95}.
One approach introduces an interpolation of the lattice gauge field to a
continuum gauge field.
Then the fermion functional integrals are regulated, for each individual
gauge field, by another regulator, which will be denoted here by~$M$.
Examples include Pauli-Villars or a finer lattice of spacing~$1/M$.
The idea is not that new~\cite{Flu82,Goe92,Goe93}---indeed, for some
time Jan Smit has described the approach as desperate~\cite{Smi87}.
But, spurred by ref.~\cite{tHo94}, the approach has recently attracted
more attention~\cite{Hsu95,Kro95,Her95,Bod95}.

With two cutoffs the first question is whether they are taken away
together,
\begin{equation}\label{limit 1}
M\to\infty,\quad a\to0,\quad Ma~{\rm fixed};
\end{equation}
or separately,
\begin{equation}\label{limit 2}
M\to\infty~{\rm with}~a~{\rm fixed},\quad a\to0~{\rm later}.
\end{equation}
Limiting procedure~(\ref{limit 1}) is close to the strategy with
fermions on the same lattice as the gauge field: even if there are two
technical regulators, there is essentially only one cutoff scale.
This is the goal that seems so difficult to reach.
On the other hand, limiting procedure~(\ref{limit 2}) is obviously
extremely costly from a practical (i.e.\ numerical) point of view:
the limit $M\to\infty$ is taken for each gauge field separately.

The ``well-known'' properties of chiral fermions are usually
established in perturbation theory by a procedure~(\ref{limit 2}).
One introduces a cutoff~$M$ for the fermions, presuming gauge fields
to be smooth and weak on the scale~$M$.
Only after taking $M\to\infty$ does one contemplate a regulator for the
gauge field.
Thus, one might expect a two-cutoff scheme starting with interpolated
lattice gauge fields to work, even if it's expensive, provided the
details are worked out~\cite{tHo94}.

Be that as it may, with two cutoffs coupling-constant renormalization is
not carried out in a satisfactory way.
In a nonperturbative renormalization one would take $M\to\infty$,
holding physical quantities (or a fiducial physical quantity)
constant.
But physics comes after integrating over all gauge fields, which
contradicts the notion of taking the limit gauge field by gauge field.
The alternative is a perturbative renormalization, introducing, say, a
partially renormalized coupling
\begin{equation}
g_0^{-2}(a) = g_{00}^{-2}(a;M) + \beta_f \log Ma
\end{equation}
where the coefficient~$\beta_f$ comes from one fermion loop, and
holding~$g_0^2(a)$ constant as $M\to\infty$.
Such a set-up is not {\em non}perturbative in the same sense as
vector-like lattice gauge theories.

The aim of this talk
is to characterize terms in the fermion effective action that are
suppressed by $1/M^{2n}$.
In a vector-like theory these terms are not very interesting, because
their effect on physical observables is (after renormalization) nil.
In a chiral theory, however, they can break gauge symmetry---because the
regulator always breaks chiral gauge symmetry.

My original goal was to control the cutoff artifacts to the extent that
it would be permissible (for some specific regulator(s)) to implement
limiting procedure~(\ref{limit 1}).
This has not (yet) proven possible.
The analysis does, however, suggest a kind of Symanzik improvement to
limiting procedure~(\ref{limit 2}).
This might prove valuable to a desperate numerical simulation.
Furthermore, many properties of the cutoff artifacts are common to
a wide class of regulators, and it might be fruitful to check how the
gauge-breaking terms can affect physical quantities, for example in
perturbation theory.

\section{INTERPOLATION}
There are several properties that one would like from an interpolation.
First, one would like the only singularities in the field
strength~$F_{\mu\nu}(x)$ to be instanton-like directional
singularities.
Second, the interpolation scheme should also generate an
interpolation~$g(x)$ of lattice gauge transformations, so that the gauge
potential transforms as
${}^gA_\mu(x) = g(x)(\partial_\mu + A_\mu(x))g^{-1}(x)$.
Third, a composition of lattice gauge transformations should be
interpolated as
\begin{equation}
\begin{array}{l}
g(s)    = g_2(s)           g_1(s) \quad \Rightarrow \\
\qquad g(x; U) = g_2(x;{}^{g_1}U) g_1(x; U),
\end{array}
\end{equation}
where the dependence of the interpolation on the underlying lattice
gauge field is emphasized.
Finally, the interpolation ought to provide an explicit form
for~$A_\mu(x)$.
It is needed for computing of eigenvalues (in a Pauli-Villars scheme) or
for constructing a finer-grained lattice gauge field.

The interpolation given in ref.~\cite{Goe93} satisfies these
requirements.

\section{GENERAL FEATURES}
The fermion Boltzmann factor is given (formally) by
\begin{equation}\label{formal}
e^{-\Gamma(A)}=\int\cD\psi_+\cD\psi_+^\dagger\,
e^{-S(A,\psi_+,\psi_+^\dagger)}
\end{equation}
where the action $S=\int d^4x\,\psi_+^\dagger\slsh{D}_+\psi_+$.
Since~$\psi_+$ is an element of a positive-chirality vector space, but
$\slsh{D}_+\psi_+$ is an element of a negative-chirality vector space,
$e^{-\Gamma}$ is not a determinant.
Nevertheless, from eq.~(\ref{formal}) one can abstract the feature of
fermion-number nonconservation~\cite{tHo76}:
if the vector-like Dirac operator $\slsh{D}$ has positive (negative)
chirality zero modes, the Boltzmann factor vanishes; inserting, however,
enough factors of $\psi_+$ ($\psi_+^\dagger$) yields a nonvanishing
integral.
One would like to preserve this property in the regulated theory.
One would also like to find
\begin{enumerate}
\item
$\real\Gamma({}^gA)=\real\Gamma(A)$,
\item
$\imag\Gamma({}^gA)=\imag\Gamma(A)$ in an anomaly-free representation,
\item
$\imag\Gamma(A)\neq0$ in a complex representation.
\end{enumerate}

\section{A DEFINITION OF $\Gamma$}
One way to define a regulated fermion Boltzmann factor starts by
considering the operator
\begin{equation}
\hat{D}=
\slsh{\partial} + \half(1-\gamma_5)\slsh{A}.
\end{equation}
The action is now $S=\int d^4x\,\bar{\psi}\hat{D}\psi$, and~$\hat{D}$
has different left and right eigenvectors.
If $\lambda_n\neq0$
\begin{equation}
i\hat{D}\eta_n=\lambda_n\eta_n, \quad
(i\hat{D})^\dagger\chi_n=\lambda_n^*\chi_n.
\end{equation}
In addition, $\hat{D}$ ($\hat{D}^\dagger$) annihilates the positive-
(negative-)chirality zero modes of $\slsh{D}$, denoted $\varphi_\pm$.
The functional integral over~$\psi$ and~$\bar{\psi}$ is defined through
the eigenmodes $\eta_n$, $\chi_m$, and~$\varphi_\pm$.
After integrating over the nonzero modes one obtains the Boltzmann
factor
\begin{equation}\label{tilde Gamma}
e^{-\tilde{\Gamma}}=\prod_{n:\lambda_n\neq0}(-i\lambda_n)
=\prod_{n:\real\lambda_n>0}\lambda_n^2
\end{equation}
Integration over zero modes (when needed) yields zero, unless enough
powers of~$\psi$ or~$\bar{\psi}$ are inserted.
But even then $\tilde{\Gamma}$ must be computed.

  From eq.~(\ref{tilde Gamma}), one posits a class of regulated
effective actions is given by
\begin{equation}\label{D hat Gamma}
\tilde{\Gamma}_L=-\half\sum_n^\infty L(\lambda_n^2/M^2)
\end{equation}
where the cutoff function~$L$ satisfies
\begin{equation}
\lim_{x\to0}\frac{L(x)}{\log x}=1,
\end{equation}
\begin{equation}
L(\infty)= L'(\infty)= L''(\infty)=\cdots=0.
\end{equation}
Examples include Fujikawa's regulator~\cite{Fuj79} and 't~Hooft's
Pauli-Villars regulator~\cite{tHo94}.

As with all regulators of chiral gauge theories, the ones given in
eq.~(\ref{D hat Gamma}) break gauge symmetry.
After a gauge transformation $g=e^\omega$
the eigenvalues of $\hat{D}$ change.
As a consequence, neither $\real\tilde{\Gamma}$ nor
$\imag\tilde{\Gamma}$ is gauge invariant.

If the regulator function~$L(x)$ is smooth and analytic everywhere, the
manipulations of the appendices of ref.~\cite{Kro95} show that the gauge
variation can be organized by powers of $M^2$.
One finds
\begin{equation}\label{expansion}
\delta_\omega\tilde{\Gamma}_L= \sum_{n=-1}^\infty M^{-2n} f_L^{(n)}(0)
\int d^4x\,\omega^a \alpha^a_{4+2n},
\end{equation}
where $f_L(x)=xL'(x)$ and the superscript on~$f_L$ denotes
differentiation with respect to~$x$.
(And $f^{(-1)}(x)=-\int_x^\infty dy\,f_L(y)$ is the first
anti-derivative.)
The gauge field comes in through the~$\alpha^a_d$, described in a bit
more detail below.

The utility of eq.~(\ref{expansion}) is that it disentangles the
$L$-dependent coefficients~$f_L^{(n)}(0)$ from the
operators~$\alpha^a_{4+2n}$, which are the same for all cutoffs in the
class defined by eq.~(\ref{D hat Gamma}).
Thus, for all these cutoffs the dimension-four ($n=0$) operator is
multiplied by the universal~$f_L(0)=1$, but the $M$-dependent terms are
multiplied by nonuniversal derivatives.
Moreover, an operator that causes difficulty in one cutoff will cause
the same difficulty for others, unless by design or good fortune the
appropriate derivative~$f_L^{(n)}(0)$ vanishes.

Detailed expressions for the operators~$\alpha^a_d$ are too long
to present here.
They are traces of the $d$-dimensional combination of~$\hat{D}$
and~$D_\mu$ yielding functionals of~$A$, rather than a differential
operators.
After the (gauge-group and Dirac-matrix) trace one can write
$\alpha^a_d = \alpha^a_{dR} + i \alpha^a_{dI}$.
In this notation the first several imaginary parts can be identified in
a less obtuse way: the two-dimensional term~$\alpha^a_{2I}=0$, and the
four-dimensional term~$\alpha^a_{4I}$ is the consistent
anomaly~\cite{Bar69,Wes71}.
Before the trace the expression that generates~$\alpha^a_{6I}$ would
generate the consistent anomaly of a six-dimensional theory; since
Dirac-matrix traces vary somewhat with dimension, after the trace
the expression in four dimensions differs in detail.

The real parts~$\alpha^a_{dR}$ are well-known.
In general they can be related to local counter-terms~\cite{Bar69}
$\int d^4x\,\omega^a\alpha^a_d=\delta_\omega s_d$,
where the $s_d$ are local (gauge noninvariant) interactions.
Since the first two do not vanish they must be added explicitly
to obtain an effective action
\begin{equation}
\Gamma=\tilde{\Gamma}-f^{(-1)}(0)M^2s_2-s_4
\end{equation}
that is gauge invariant after the taking~$M\to\infty$.

\section{ANOTHER DEFINITION}
Another approach deals with eigenvalues of vector-like Dirac operators
only.
It would take
\begin{equation}\label{Gamma eta}
e^{-\Gamma}=\sqrt{{\det}_{L_1}\slsh{D}}\,\exp(i\eta_{L_2}),
\end{equation}
where ${\det}_L\slsh{D}$ is the regulated Boltzmann factor of a
vector-like theory, and $\eta_L$ is a certain (UV-regulated) property a
a five-dimensional Dirac operator~\cite{Alv86}.
After Pauli-Villars regulators (for example) are removed $\eta_L$
provides the anomalous Wess-Zumino action~\cite{Wes71} and
gauge-invariant (but parity-violating) terms.

The work of G\"ockeler and Schierholz~\cite{Goe92}, as well as some more
recent papers~\cite{Hsu95,Bod95}, introduces a Boltzmann factor with the
same properties as eq.~(\ref{Gamma eta}) under gauge transformations.
These~$\real\Gamma$ are all gauge invariant, so their cutoff effects are
neither harmful nor interesting.
On the other hand, at finite~$M$ the imaginary
part~$\delta_\omega\imag\Gamma\neq0$, even when anomalies cancel.
But, as long as the regulator used for~$\eta_L$ reproduces the
{\em consistent\/} anomaly, it seems likely that the higher-dimension
analogues are the same as~$\alpha_{6I}$, etc, described in sect.~4.

\section{REDUCTION OF CUTOFF EFFECTS}
If there are higher-dimension gauge-breaking terms in~$\real\Gamma$ it
is essential to remove them.
Since they are equivalent to local counter-terms, they will mix with
lower-dimensional terms, leaving the interacting theory with
gauge-breaking of~$\Order(1/(aM)^{2n}$.
The first several such terms can be eliminated by choosing the cutoff
function so that successive derivatives $f^{(n)}(0)=0$.
It does not seem possible to eliminate all cutoff effects
by setting all $f^{(m)}(0)=0$.
(A function like that has an essential singularity at $x=0$.
But the manipulations in the appendices of ref.~\cite{Kro95} require
Taylor expansions of $f(x)$ too close the origin to remain valid in this
case.)
Barring a more robust proof, the most optimistic prospect is that an
improved cutoff would accelerate the convergence to the limit
$M\to\infty$.

If the gauge breaking comes only from~$\imag\Gamma$ the conclusions
are less obvious, because one cannot express these terms as
local functionals of~$A_\mu$ and its derivatives.
For example, the anomalous part is given by the Wess-Zumino action,
an integral over a five-dimensional manifold.
If the conclusion of the mixing argument applies in this setting, then
the higher-dimensional terms would filter back down to the anomaly, the
imaginary part of lowest dimension.
It would then seem that an anomaly-free theory contains no residue of
the cutoff's gauge breaking.
That seems too good to be true, but it could be examined by inserting
$\alpha_{6I}$ into Feynman diagrams.

\section{CONCLUSIONS}
This talk
has exposed some features of the cutoff effects in chiral gauge theories
with two regulators, one for the gauge fields and another for the
fermions.
In particular, there is a connection between the gauge-breaking cutoff
artifacts of the imaginary part and expressions that lead to anomalies
in higher-dimensional theories.
Inasmuch as the same terms arise in a variety of different regulators,
this may be a first step towards deriving some general properties.

For example, there is some disagreement~\cite{Gol95} whether
the overlap formalism~\cite{Nar95} yields chiral fermions.
In this formalism the imaginary part of the effective action breaks
gauge invariance at finite lattice spacing.
One can imagine setting up the overlap in a two-cutoff scheme, and then
examining the large $Ma$ limit, as above.
Since the dimension-four gauge breaking is the consistent anomaly, it
again seems likely that at dimension six~$\alpha^a_{6I}$
appears, the same expression as above.
Depending on how it (or other gauge breaking) trickles down to physical
quantities, one may learn whether the overlap can implement limiting
procedure~(\ref{limit 1}), or whether it too must live with the
desperate limiting procedure~(\ref{limit 2}).

Many thanks to J. Bijnens, P. Hoyer, and the rest of the Nordita staff
for a pleasant visit, during which these proceedings were written up.
Fermilab is operated by Universities Research Association, Inc.,
under contract DE-AC02-76CH03000 with the U.S. Department of Energy.

\end{document}